# NeR-VCP: A Video Content Protection Method Based on Implicit Neural Representation

Yangping Lin, Yan Ke*, Ke Niu, Jia Liu, Xiaoyuan Yang *

[1] Lin and Ke should be considered joint first author

ABSTRACT

With the popularity of video applications, the security of video content has emerged as a pressing issue that demands urgent attention. Most video content protection methods mainly rely on encryption technology, which needs to be manually designed or implemented in an experience-based manner. To address this problem, we propose an automatic encryption technique for video content protection based on implicit neural representation. We design a key-controllable module, which serves as a key for encryption and decryption. NeR-VCP first pre-distributes the key-controllable module trained by the sender to the recipients, and then uses Implicit Neural Representation (INR) with a (pre-distributed) key-controllable module to encrypt plain video as an implicit neural network, and the legal recipients uses a pre-distributed key-controllable module to decrypt this cipher neural network (the corresponding implicit neural network). Under the guidance of the key-controllable design, our method can improve the security of video content and provide a novel video encryption scheme. Moreover, using model compression techniques, this method can achieve video content protection while effectively mitigating the amount of encrypted data transferred. We experimentally find that it has superior performance in terms of visual representation, imperceptibility to illegal users, and security from a cryptographic viewpoint.

I. INTRODUCTION

**Background:** With the rapid development of the mobile Internet and multimedia applications, the video transmission and communication have become very convenient. A large amount of video information about individuals, enterprises, and governments needs to be processed, transmitted, and accessed through the Internet. Due to the openness and sharing of the Internet, people can access video resources anytime and anywhere through multiple social activities, such as video conferencing, live video, and video surveillance [1-9]. Privacy and content protection technology has garnered significant attention from numerous scholars [10, 11]. Therefore, this trend has raised concerns among people regarding the privacy and security of video media [12]. Currently, privacy protection methods for video content have gradually gained popularity in the field of information security. To protect videos from unauthorized access, information leakage, theft, and other threats, a security system for video information security has been proposed, namely, video encryption [13-17]. Many researchers have proposed multimedia content protection technology based on data hiding, deoxyribonucleic acid (DNA) coding, and chaotic maps [18-28] to protect the security of image and video information. These secure systems have extremely promising applications in various fields.

**Development:** Video content protection is usually combined with video compression coding. Video compression coding is one of the most important techniques in real-time applications. However, a 2-hour Moving Picture Experts Group MPEG-4-encoded video is approximately 1.6 GB in size and needs to be transmitted and played in real-time or at high frame rates. If the encryption method is not incorporated into the encoding process, there will be a considerable additional computational burden. Traditional video encryption algorithms encrypt the entire video stream such as Data Encryption Standard (DES), RC4, Simplified Data Encryption Standard (SDES), and Modified Advance Encryption Standard (MAES) [29-31]. This type of encryption can have acceptable high-level security; nonetheless, it does not use the characteristics of the video, resulting in too low encryption efficiency. Another interesting topic is video selective encryption techniques based on H.264 and HEVC video compression. By analysing the encoded video stream, this selective encryption method encrypts important video elements to reduce complexity and maintain certain security [13, 15, 16, 20, 25, 27, 29-34]. Also, some scholars use reversible watermarking technology for video copyright protection and ownership identification, and authorized users can recover video losslessly through watermarking [19]. However, few studies have protected video information security without modern cryptography and chaotic encryption, and these experience-based video encryption methods need to encrypt the specifically encoded video or protect the watermark in manually designed manners, which is difficult to apply to various video compression standards.

**Deep learning** has brought new changes to the field of information security, some deep learning-based methods implement adaptive content protection and demonstrate the exciting secure performance [35]. Recently, implicit neural representation (INR) has been a hot topic of deep learning. Considering its excellent performance in video representation and adaptability, it has been proposed to replace part of the video encryption algorithm to achieve adaptive video encryption for content protection.

In this work, we propose NeR-VCP, a novel video content protection network based on implicit neural representation. The proposed network, which is inspired by video neural representation, encrypts a plain video into a cipher implicit neural network using a key-controllable module as the key. The key-controllable module trained by the sender is pre-distributed to the receiver. The receiver can decrypt the video by the pre-distributed key-controllable module. Therefore, the key-controllable module guarantees that authorized users can decrypt the video, prevent unauthorized access, and we can compress the video using model compression methods to reduce the transmission cost of cipher neural networks. In addition, without multiple transmissions of the parameters in key-controllable module, the sender can achieve content protection of different videos through the pre-distributed key-controllable module. This means that we do not need to repeatedly train the key-controllable module for different video content.

In the experiment, the efficiency of the proposed video content protection method is evaluated using the "Big Buck Bunny" and UVG datasets, and the video representation results under different model compression techniques are compared to that of other video neural representation methods. From the results of security and pixel correlations, our method is demonstrated to have good visual security while ensuring representation efficiency compared to other encryption methods.

The remainder of this paper is structured as follows. Section II presents the related work on video content protection and INR. Regarding our video content protection method, Section III includes an overview, a training loss, a key-controllable scheme, and a neural representation for video . Experiments on our proposed method in terms of Key-controllable security analysis and video protection efficiency are presented in Section IV, which summarizes this paper.

II. RELATED WORK

Video content protection plays an important role in video surveillance, video conferences, and real-time applications, and video encryption can be categorized into two types, naïve encryption and selective encryption. The first type uses encryption algorithms to encrypt all the frames, parameters, or data contained in a video stream in sensitive areas. These encryption methods usually encrypt entire video data to protect the security of private information [3, 36, 37]. Chandrasekaran et al. [37] proposed a naive video encryption method to encrypt the entire video stream efficiently through the Henon map. This type of encryption is expensive in terms of both time and space, and bring a large amount of computational cost. Moreover, some studies can achieve the same security by encryption of selectively chosen frames or a part of video data, which was proposed by Maples et al [29]. Considering the importance of the intra-prediction model and motion vector, some studies [4-6] encrypt or scramble them to achieve video encryption. Additionally, chaotic systems [16, 17, 27-28, 30, 38], DNA coding [21-24], and quantum techniques [39] have been applied to image and video encryption algorithms. Chaotic systems, which originate from the Lorenz system, are a system of ordinary differential equations. Chaotic systems are used as pseudo random number generators for key generation and a part of encryption algorithms. A hyper-chaotic system is an extension of two or more positive Lyapunov exponents [21, 30]. Sharma et al. [30] proposed a novel secure watermark embedding technique based on graph-based transform, singular valued decomposition, and hyperchaotic encryption. The encryption methods based on DNA coding hide data by DNA Sequences and use DNA components with plaintext in a one-way function to generate ciphertext, which is similar to public key systems. Karmakar et al. [21] proposed an effective hyperchaotic DNA coding-based encryption mechanism combined with a sparse representation-based spatial video compression. These selective encryption methods not only need to compress the video before scrambling and encrypting but also can only be applied to the fixed video compression standard. Traditional video encryption scheme can better protect private information, but there are problems of difficulty in manual design and low adaptability. Moreover, these video content protection methods fail to achieve video compression after the naïve encryption of the entire video content. In addition to the encryption algorithm and data hiding are known only to the authorized sender and receiver and are also a more secure video content protection method where ciphertext attracts the adversary to decode the information. Authorized users can losslessly recover the original video via data hiding and watermarking [18-20, 40].

Currently, due to simplicity, compactness, and efficiency of implicit neural representation, it has become increasingly popular in computer vision. Generally, this neural representation method can be applied in two visual categories. The first category is pixel-wise representation, which takes pixel coordinates as inputs. In the applications of image regression, image reconstruction, and 3D view generation, several pixel-wise representation methods have been proposed [41-44]. The second category is image-wise representation which takes a time index as input. Compared to the first category, this operation can improve the speed of encoding and decoding. Neural representations for videos (NeRV) takes a timestamp as input and outputs the entire frame [45]. Recent studies [46,47] have improved the design of the architecture for better visual performance. Chen et al. [46] proposed a neural visual representation based on content-adaptive embedding and generalized it to skipped images for a given visual input. Li et al. [47] designed a novel image-wise video INR, which expedites NeRV by decomposing the image-wise method into separate spatial and temporal contexts. In the field of information security, Luo et al. [48] proposed a method to protect the copyright of NeRF models, which achieved good results in secure communication.

III. NeR-VCP METHODOLOGY

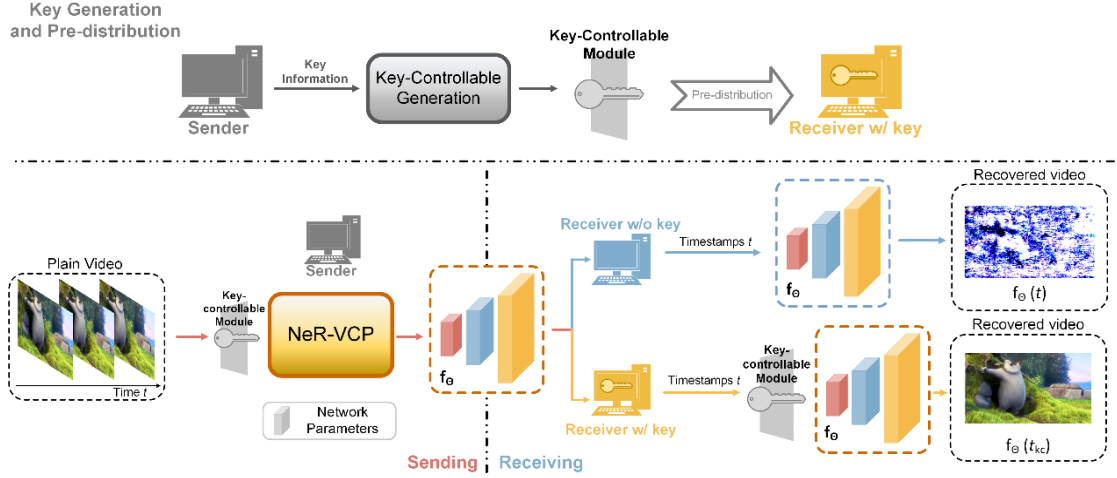

FIGURE 1 Overview of our proposed method. The upper part shows the generation and pre-distribution of key-controllable module. The lower left part represents plain video as a cipher neural network, taking a frame timestamp as input of key-controllable module and outputting the corresponding neural network. The lower right part consists of authorized and unauthorized access to video.

To achieve video neural representation while guaranteeing the imperceptibility of plain video, we construct a video content protection network based on implicit neural representation. As shown in Fig. 1, this section presents the details of the proposed model, which contains two main modules: a key-controllable module and a video neural representation module. First, the sender pre-distributes a key-controllable module to a legal receiver, which achieves access control of video content. The key-controllable module enhances the security of video content, which is regarded as a key. The video representation module securely represents the plain video, encrypts it into a neural network, and ensures the secure reconstruction of plain video via a pre-distributed key-controllable module. After receiving a cipher video representation model, which is viewed as the parameter of a unified neural network, we need to take the input embedding generated by key-controllable module as the key to decrypt the encrypted video.

A. Secure Positional Encoding Based on Key-controllable Scheme

As another implicit neural representation, pixel-wise visual representations [49] take spatial-temporal coordinates (pixel index) as inputs. In other words, pixel-wise representations output the RGB value for each pixel from the pixel index, while NeRV outputs a whole image. In NeRV, for each video, the frame index is a key factor in the implicit visual representation of NeRV. Each video $V = \{v_t\}_{t=1}^{T} \in \mathbb{R}^{T \times H \times W \times 3}$ is represented by a function: $f_\theta : \mathbb{R} \to \mathbb{R}^{H \times W \times 3}$, where the input is a frame index (timestamp) $t$ and the output is the corresponding frame in video $v_t \in \mathbb{R}^{H \times W \times 3}$. A neural network $\theta$ is used to fit the coding function $v_t = f_\theta(t)$. Video encoding is realized by the neural network $f_\theta$ corresponding to this function, and the corresponding RGB image can be obtained as long as the corresponding timestamp is input to $f_\theta$.

Although deep neural networks can now achieve many complex tasks in many fields, as a tool for universal function approximators [50], if we train the timestamp $t$ directly as an input to the neural representation network, this can lead to poor results, as mentioned in [49, 51]. Mapping low-dimensional timestamps to high-dimensional spaces will make the neural network more conducive to fitting video data with high-frequency information. To better represent the positional information of the frame, Positional Encoding [52], which is a method of quadratic representation of the positional information of the elements for each element in the sequence, was introduced in transformer as follows:

$$\Gamma(t) = \left(\sin\left(b^0 \pi t\right), \cos\left(b^0 \pi t\right), \ldots, \sin\left(b^{l-1} \pi t\right), \cos\left(b^{l-1} \pi t\right)\right) \quad (1)$$

where $b$ and $l$ are hyperparameters of the networks. Taking a timestamp $t$ as input to the PE function $\Gamma(\cdot)$, normalized between $(0,1]$, the output of the PE function serves as input embedding to the following neural network.

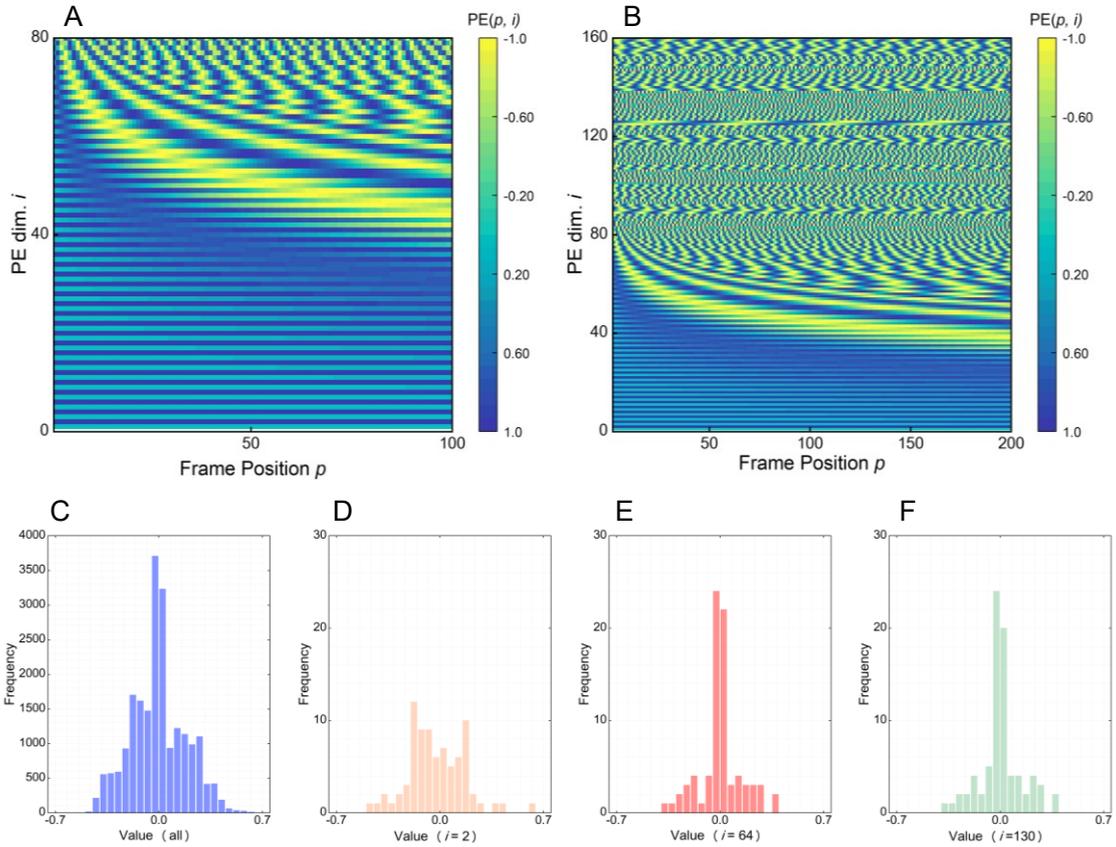

FIGURE 2 Histogram of the frequency distribution of key-controllable PE. A and B visualize the embedding features of PE function with dims of 80 and 160 and frame index of 100 and 200. C-F. the output features distribution with different frame indexes.

The PE function uses a simple yet genius technique to model positional information [53]. Nevertheless, such simple embedding of PE lacks data security and specificity. Furthermore, considering the practicability of our scheme, to improve the security of our scheme, we refer to previous work [54] that added a key-controllable scheme to PE function to enable the legal receiver to decrypt plain videos from the cipher neural networks through pre-distributed keys. The key-controllable design is used to handle the case where there is a specific requirement for secure transmission. The design of the key-controllable module is shown in Fig. 3. Given the timestamp $t$ of the corresponding frame $v'_t$ in a plain video $V_p = \{v'_t\}_{t=1}^{T} \in \mathbb{R}^{T \times H \times W \times 3}$, a key mask is generated by a key-controllable encoder, which is composed of fully connected (FC) layers. The timestamp $t$ is then fed into the first FC layer to map into vector space $\mathbb{R}^{2l \times 1}$, where $2l$ is the channel of the output of the PE function. Then, we use the latter two layers of fully connected networks to fine-tune the key embedding while fitting the data distribution of the PE function and increasing the data adaptability. Obviously, this coding method will affect the visual performance of the video representation module. We provide the architecture details in FIG 3. In the key-controllable encoder, given the timestamp $t$ of the plain video, we apply the first two FC layers with an output feature to $2l$ following GELU activation; then, we multiply the input embedding of PE function by key mask to obtain the key embedding.

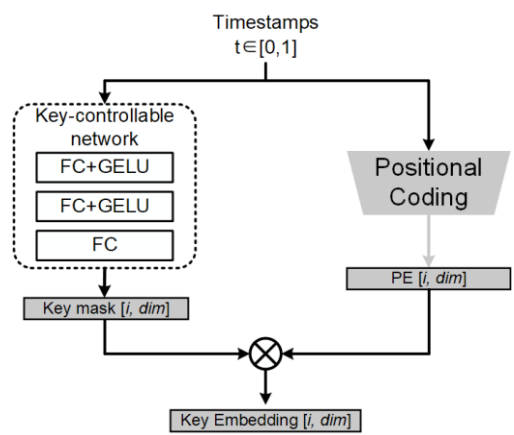

FIGURE 3. The architecture of the proposed key-controllable module. "FC+GELU" represents the FC layers and GELU activation.

B. Neural representations for videos

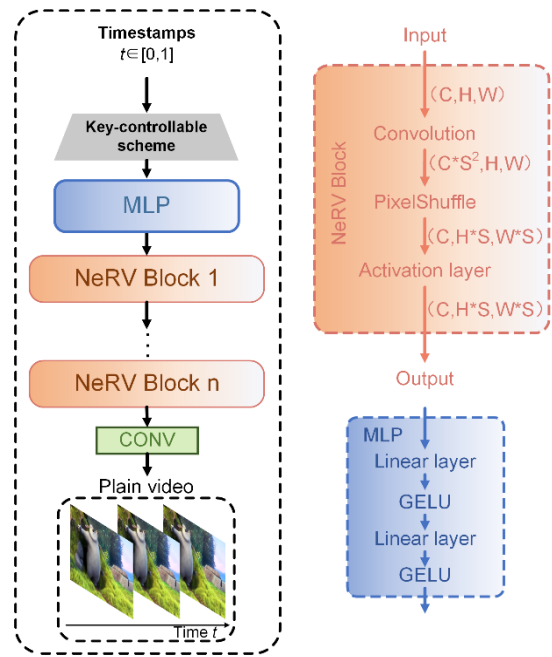

FIGURE 4. Framework of NeR-VCP. Blue: MLP block. Orange: Neural representations for videos

(NeRV) block. Black: NeR-VCP consists of Key-controllable scheme, an MLP, and NeRV blocks.

NeRV is composed of MLP and several NeRV blocks for recovering resolution, given the timestamp embedding, which outputs the corresponding RGB frame in a secret video, as shown in FIG 4. When the video resolutions are large, using MLPs to directly generate all pixel values of the frames can cause low efficiency and high computing costs. Therefore, NeRV takes several NeRV blocks following the MLP layers, and while reducing the decoding time, convolutional kernels can also be shared in the pixel-wise space. Inspired by the method of super-resolution network [55], NeRV adopts the PixelShuffle technique for upscaling tasks in the NeRV block. The other methods for upscaling a low-resolution image include a deconvolution layer and convolution with a fractional stride of $\frac{1}{r}$. These methods increase the computational cost by a factor of $r^2$, since upscaling convolution operations occur in high-resolution space. NeRV introduces an effective method named sub-pixel convolution to implement the upscaling operation:

$$\mathbf{I}^{SR} = f^L\left(\mathbf{I}^{LR}\right) = \mathcal{PS}\left(W_L * f^{L-1}\left(\mathbf{I}^{LR}\right) + b_L\right), \text{ when } \mathrm{mod}(k_s, r) = 0 \qquad (2)$$

where $\mathcal{PS}$ is the pixel-wise shuffling operator that adjusts the tensor with a size of $H \times W \times C \times r^2$ to the elements of a $rH \times rW \times C$ tensor, $r$ is the upscaling factor. $k_s$ is the size of the filter in the convolution. The convolution operator $W_L$ has the shape of $n_{L-1} \times r^2 C \times k_L \times k_L$. This operation is described as PixelShuffle in torch, as shown in Fig. 5:

$$\mathcal{PS}(T)_{x,y,c} = T_{\lfloor x/r \rfloor, \lfloor y/r \rfloor, C \cdot r \cdot \mathrm{mod}(y,r) + C \cdot \mathrm{mod}(x,r) + c} \qquad (3)$$

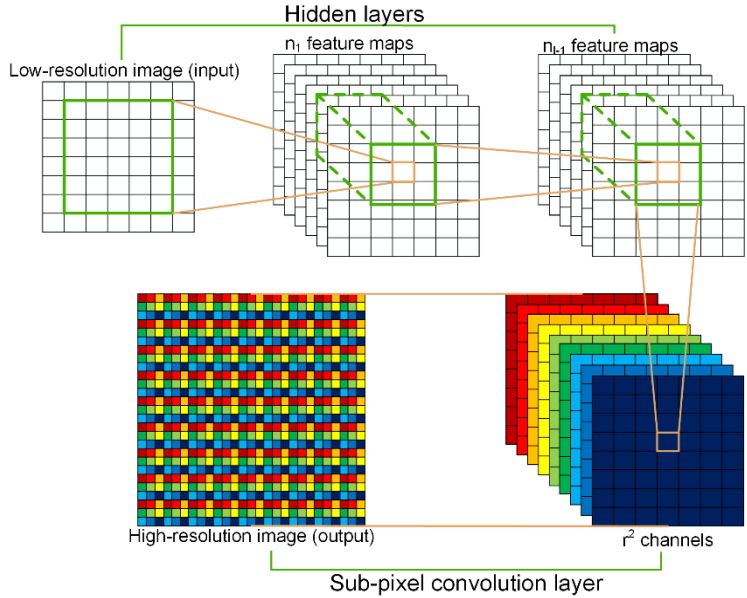

FIGURE 5. The sub-pixel convolutional neural network (PixelShuffle). The hidden layers represent the features of convolution layers. The sub-pixel convolution layer represents PixelShuffle operation.

The architecture of NeRV network is shown in Table 1. The MLP block applies two linear layers followed by a GELU activation layer. The NeRV architecture consists of 5 NeRV blocks, with upscaling factors of 5, 3, 2, 2, and 2 for 1080p videos, and 5, 2, 2, 2, and 2 for 720p videos. NeRV block consists of a 3×3 convolution layer with a stride of 1, PixelShuffle operation, and GELU activation. After the last NeRV block, we use a 1×1 2Dconvolution layer with stride of 1 and output channel of 3. We can train NeRV network with different model sizes by adjusting the values of $C_1$ and $C_2$.

Table 1. NeRV architecture for two types of videos. The values of values of $C_1$ and $C_2$ are adjusted to obtain different model sizes.

| Layer | Modules | 720p | | 1080p | |
|---|---|---|---|---|---|
| | | Upscale Factor | Output Size | Upscale Factor | Output Size |
| 0 | Key-controllable module | - | $2l \times 1$ | - | $2l \times 1$ |
| 1 | MLP&reshape | - | $C_1 \times 16 \times 9$ | - | $C_1 \times 16 \times 9$ |
| 2 | NeRV block | 5× | $C_2 \times 80 \times 45$ | 5× | $C_2 \times 80 \times 45$ |
| 3 | NeRV block | 2× | $C_2/2 \times 160 \times 90$ | 3× | $C_2/2 \times 240 \times 135$ |
| 4 | NeRV block | 2× | $C_2/4 \times 320 \times 180$ | 2× | $C_2/4 \times 480 \times 270$ |
| 5 | NeRV block | 2× | $C_2/8 \times 640 \times 360$ | 2× | $C_2/8 \times 960 \times 540$ |
| 6 | NeRV block | 2× | $C_2/16 \times 1280 \times 720$ | 2× | $C_2/16 \times 1920 \times 1080$ |
| 7 | Output layer | - | $3 \times 1280 \times 720$ | - | $3 \times 1920 \times 1080$ |

C. Loss

In the entire model training process, we first use the MSE loss function to train the key-controllable module, which optimizes the loss of key mask and input embedding of PE function, as follows:

$$L_{KcPE} = MSE(M_{key}, E_{pe}) \qquad (4)$$

where $MSE$ denotes the MSE loss, $M_{key}$ and $E_{pe}$ are the key mask and input embedding, respectively. In the first training process, we take the timestamps as the inputs of the key-controllable encoding and PE function.

In the NeRV training process, we define the loss function as a composite loss of SSIM and L1, which optimizes the loss of the predicted image and the ground-truth image at pixel-wise locations as follows:

$$L = \frac{1}{T}\sum_{t=1}^{T} \alpha \|f_\theta(t) - v_t\|_1 + (1-\alpha)\left(1 - \text{SSIM}(f_\theta(t), v_t)\right) \qquad (5)$$

Where $T$ is the frame timestamp, $f_\theta(t) \in \mathbb{R}^{H \times W \times 3}$ is the predicted frame of NeRV, $v_t \in \mathbb{R}^{H \times W \times 3}$ is the ground truth, $\alpha$ is a hyperparameter for making a trade-off between the L1 and SSIM loss component, $\alpha$ is set to 0.7.

III. EXPERIMENTS AND ANALYSIS

For comparison with other multimedia encryption methods in terms of key-controllable security analysis and video protection efficiency. We evaluated the validity of the key-controllable module from analysis of imperceptibility, key-controllable designs, noise attacks, and pixel correlation. Additionally, video representation efficiency serves as a crucial metric in evaluating the performance of time and space costs for encrypting video.

A. Implementation Details

We adopt the scikit-video dataset "Big Buck Bunny" sequence, which has 132 frames of 720p resolution. Furthermore, we conducted experiments on UVG [56], which consists of 3900 frames from seven 1080p videos in total, to compare the proposed model with SOTA video compression methods for video protection efficiency. For imperceptibility and design rationality, we analysed and compared the visual performance of the recovered encrypted videos from NeR-VCP with different key-controllable designs. For the pixel correlation and security of the proposed model, we tested the secure performance with different sizes of NeR-VCP compared to other encryption methods. For the efficiency of the proposed model, we tested the training speed, decoding FPS, MS-SSIM, and PSNR. To assess the visual quality of our method, we have employed MS-SSIM and PSNR metrics.

Our model was trained with Python based on PyTorch. GeForce RTX 4060 GPU and CUDA 12.1 are used, and we train the model with full precision (FP32). Several hyperparameters need to be illustrated, in addition to the model parameters obtained by training. We chose the Adam optimizer [57] as the optimization method and set a learning rate of 0.0005. For "Big Buck Bunny" dataset, unless otherwise stated, we train NeRV 2400 epochs and set the batch size to 1 in experiments of representation efficiency and compression. For UVG dataset, the model was run for 300 epochs, with a batchsize of 2. For the training process of the key-controllable module, we trained it with Adam optimizer, 30 epochs, and a learning rate of 0.0001.

## B. Key-controllable Security Analysis

To validate the rationality of our proposed key-controllable module's design and assess its practical effectiveness, we conducted experiments of imperceptibility, key-controllable performance, and noise attack. These experiments allowed us to comprehensively compare the visual performance and security of the video content. Furthermore, we ensured the stability and reliability of our method in real-world scenarios, thus strengthening the credibility of our findings.

**Evaluation on imperceptibility.** The imperceptibility of plain videos is the most critical performance evaluation index of a video content protection system. Generally, we expect that the encryption operation will not cause model parameters to reveal video content. NeR-VCP consists of two parts: a key-controllable module and a video representation module. Generally, as the key data representing the plain video, the structure and parameters of the corresponding video representation module need to be protected. The sender can use a pre-distributed key-controllable module, it was transmitted through the secret channel before the encryption of the plain video. As a tool for representing a plain video, a video representation module has a large number of parameters and needs to be transmitted via a public channel. Normally, the adversary will not know our proposed model framework. To measure the imperceptibility of the model under extreme conditions, we assume that the structure and parameters of our video representation module, but not the relevant weight and parameters of the key-controllable module, are obtained. We remove the key-controllable scheme in the key-controllable module, leaving only the PE function, which generates the timestamp embedding as input to the video representation module. We tested and compared the visual performance of our model with epoch of 1200 on "Big Buck Bunny" sequence, as shown in Fig. 6. According to the results, NeR-VCP visually has high imperceptibility and satisfactory security. When no key-controllable module is involved, the adversary cannot obtain video information from the video representation module, which ensures the security of the encrypted video. Moreover, in the NeR-VCP-12.5 M model, the visual expression of our model even exceeds that of ground truth (GT), showing a certain denoising ability.

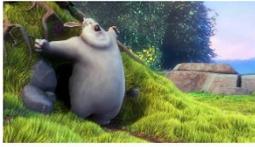

FIGURE 6. Video privacy protection visualization. w/ key represents video decryption with the key-controllable scheme.

**Evaluation on key-controllable designs.** We provide two different key-controllable schemes: learnable key-controllable encoding (LAE) and fixed key-controllable encoding (FAE). The LAE method refers to the full participation of the key-controllable module in the whole training of the model to optimize parameters. The FAE method refers to the fixed network parameters of trained FC layers and does not participate in the training of the video representation model. Specifically, the appropriate key-controllable module is trained by FAE first, and then only the video representation module is trained without fine-tuning the key-controllable module. In both the LAE and FAE, we add a key-controllable scheme to the PE function and use the same train setting. Comparing these two coding methods, the advantages of FAE are that plain videos have better security and visual performance. The advantage of FAE is that a pre-distributed key-controllable module can encrypt and protect several videos. This approach avoids multiple transmissions of key-controllable modules, ensuring both practicality and security.

We tested and compared the performance of our proposed method under two encoding methods with 1200 epochs, as shown in Table 2, we can find that FAE method with the same model size has greater visual security and better performance with high PSNR and MS-SSIM. Therefore, considering that NeR-VCP model is designed to ensure the security of the transmission process, the FAE method has a pre-distributed key-controllable network for the secure representation of multiple videos, and this method has greater confidentiality and application prospects.

Table 2. PSNR and MS-SSIM results for video content protection

| Model | Key-controllable | Our-FAE | | Our-LAE | |
|---|---|---|---|---|---|
| | | PSNR | MS-SSIM | PSNR | MS-SSIM |
| Ours-3.2 M | w/ key | **33.84** | **0.9738** | 33.30 | 0.9731 |
| | w/o key | **5.81** | **0.0889** | 7.46 | 0.0594 |
| Ours -6.4 M | w/ key | **37.47** | **0.9884** | 36.97 | 0.9881 |
| | w/o key | **6.27** | **0.0775** | 7.66 | 0.0885 |
| Ours -12.5 M | w/ key | **41.25** | **0.9953** | 40.52 | 0.9949 |
| | w/o key | 7.9 | 0.1178 | **7.89** | **0.0898** |

**Evaluation on noise attacks.** Noise attacks in multimedia fields are pixel-wise attacks, which are usually used to modify and transform the pixels themselves. According to the particularity of our proposed method, the protection of video is based on the existence of key-controllable module. In addition to comparing the differences between the model with and without the key-controllable scheme, we conducted an anti-attack experiment. In the experiment of the anti-attack capacity, we assumed that the attacker is aware of the positional encoding method, but does not obtain the details of the key-controllable module. Noise attacks based on multiple data distributions were used as a substitute for the key mask generated by key-controllable module, several types of noises were combined with the output of PE function and input into the video representation module after noise pollution.

Table 3. Experimental resutls for different noise attacks

| Method | Params. | Metrics | Gaussian | Uniform | Bernoulli | TruncN(0.05) | TruncN(0.1) | TruncN(0.15) | TruncN(0.2) | TruncN(0.25) |
|---|---|---|---|---|---|---|---|---|---|---|
| LAE | 3.2 M | PSNR | 7.36 | 10.73 | 9.44 | 18.53 | 18.24 | 17.79 | 17.23 | 16.56 |
| | | MS-SSIM | 0.044 | 0.132 | 0.0879 | 0.5509 | 0.5246 | 0.4876 | 0.4439 | 0.3958 |
| | 6.4 M | PSNR | 7.54 | 10.87 | 9.42 | 18.71 | 17.61 | 18.21 | 16.92 | 16.16 |
| | | MS-SSIM | 0.0502 | 0.1457 | 0.1082 | 0.561 | 0.4637 | 0.517 | 0.4083 | 0.3539 |
| | 12.5 M | PSNR | 7.91 | 11.59 | 10.09 | 18.77 | 18.32 | 17.82 | 17.3 | 16.73 |
| | | MS-SSIM | 0.0494 | 0.167 | 0.1191 | 0.569 | 0.522 | 0.4734 | 0.4269 | 0.3818 |
| FAE | 3.2 M | PSNR | **5.58** | **7.42** | **6.4** | **17.42** | **16.16** | **14.46** | **12.77** | **11.32** |
| | | MS-SSIM | **0.0647** | **0.0968** | **0.0826** | **0.4792** | **0.3816** | **0.2835** | **0.2066** | **0.1537** |
| | 6.4 M | PSNR | **6.04** | **7.75** | **7.02** | **17.78** | **16.34** | **14.58** | **12.87** | **11.42** |
| | | MS-SSIM | **0.0521** | **0.0972** | **0.0814** | **0.4959** | **0.3865** | **0.2828** | **0.2042** | **0.1508** |
| | 12.5 M | PSNR | **7.47** | **9.64** | **8.79** | **17.91** | **16.78** | **15.5** | **14.17** | **12.97** |
| | | MS-SSIM | **0.0709** | **0.132** | **0.1182** | **0.5017** | **0.3984** | **0.3031** | **0.2286** | **0.178** |

To investigate the responses of the proposed model under more objective and standardized conditions, we choose the Gaussian, uniform, Bernoulli, and a series of truncated normal noises with a range of [-1, 1] as noise attack methods. From these results in Table 3, we can draw the following conclusions. First, the ability to resist attack under Gaussian, uniform, and Bernoulli noise is greater than that under truncated normal noise. Moreover, we can only obtain decrypted video with poor visual performance, which is similar to the PSNR and MS-SSIM metrics obtained with the PE function only. Second, when we pre-distributed the key-controllable module obtained by FAE to train the model, the FAE performance in resisting noise attacks was better. The major reason is that our method takes full advantage of the pre-distributed key-controllable module from FAE, and only needs to consider training for the video representation module. Finally, the ability of our method to resist truncated normal noise needs to be improved, yet such weak points should not obscure the innovative point that comes from the proposed model presented in this work.

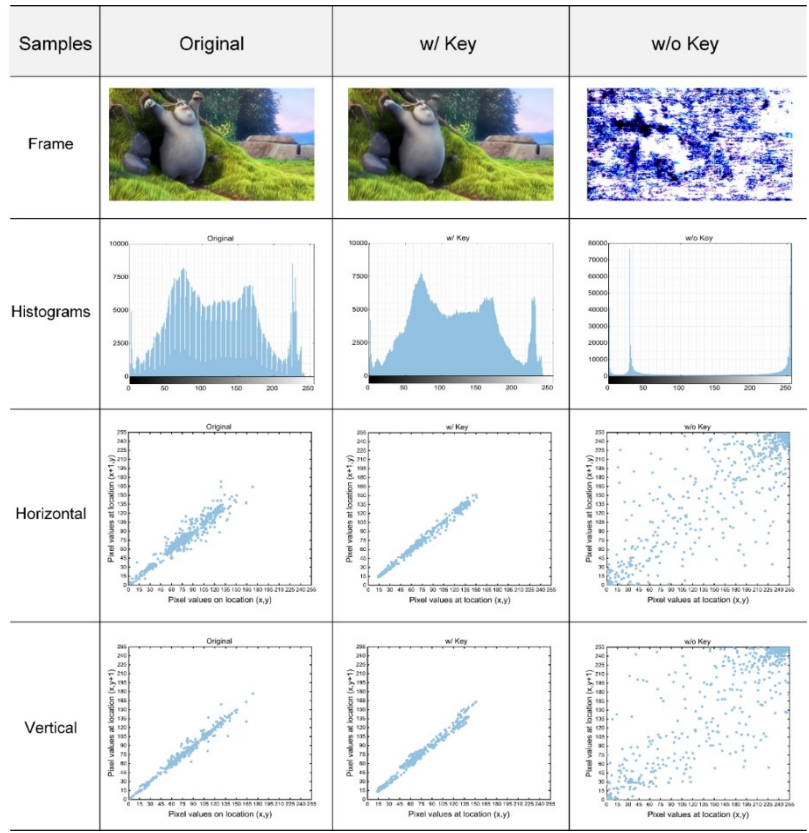

Figure 7. Histograms and horizontally and vertically adjacent pixel correlations in plain, decrypted, and encrypted frames.

**Pixel Correlation Analysis.** In regular images, pixels typically demonstrate a close correlation in all directions, constituting a crucial aspect of image structure and visual information. Nevertheless, the fundamental aim of encryption technology lies in obfuscating and eliminating this correlation among pixels, thereby safeguarding the security and confidentiality of image data. A pixel histogram directly illustrates the statistical characteristics of the frame pixels. Generally, the pixel values of the plain image conform to a specific statistical distribution, while statistical distributions of encrypted images do not have practical meaning. In this section, we used our method with a parameter size of 3.2M and an epoch of 300 as the experimental model. The second-row images in Fig. 7 show the histograms of plain, decrypted, and encrypted frames obtained by our method. The third and fourth rows in Fig. 7 show the correlation distributions of the 600 pairs of adjacent pixels randomly selected from horizontal and vertical directions of plain, decrypted, and encrypted frames. This result suggests that the histograms and pixel correlations of plain and decrypted frames have similar statistical characteristics and that decrypted frames have good visual performance. The encrypted frame is visually unable to present relevant information, and its pixel correlations are completely different from those of plain frame. There is no valuable information in our encrypted frame through human observation, and other encrypted frames of other image encryption methods [58-62] are also shown in Fig. 8.

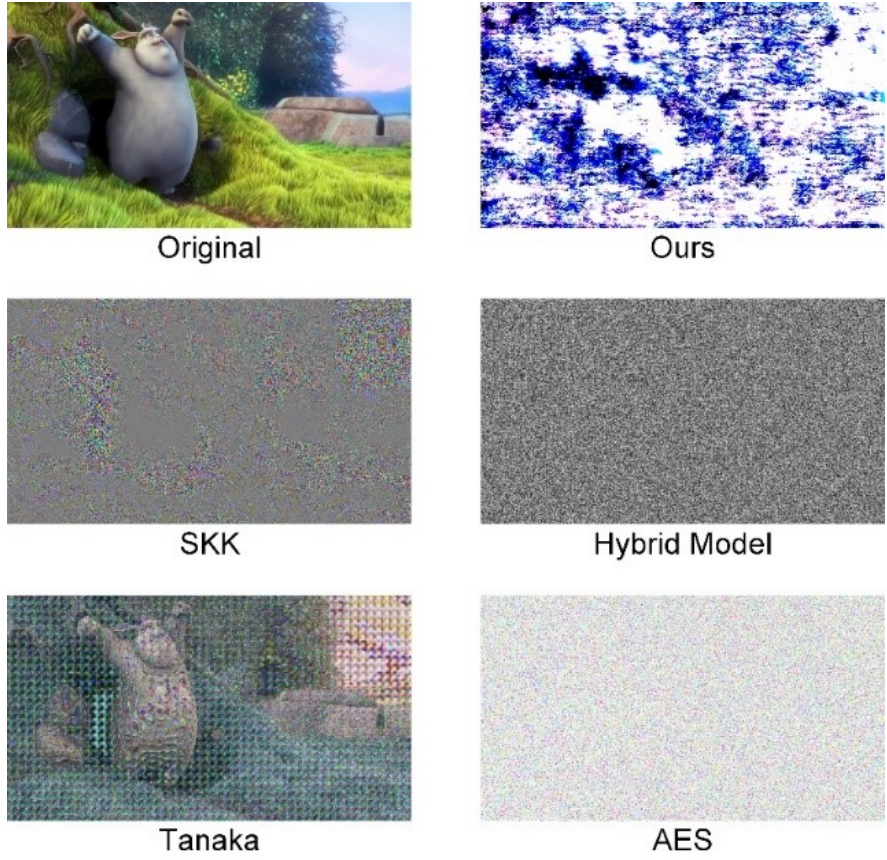

Figure 8. Examples of plain and ciphered frames by different methods.

Digital images usually present strong pixel correlations in different directions. For comparison with other encryption methods, we transform the color frame to gray one. To calculate the vertical correlation coefficient of two adjacent pixels, we have:

$$CC_v = \frac{\sum_{i=1}^{H-1}\sum_{j=1}^{W}(C_{(i,j)}-\bar{C})(C_{(i+1,j)}-\bar{C})}{\sqrt{\sum_{i=1}^{H-1}\sum_{j=1}^{W}(C_{(i,j)}-\bar{C})^2 \sum_{i=1}^{H-1}\sum_{j=1}^{W}(C_{(i+1,j)}-\bar{C})^2}} \quad (6)$$

where $H$ and $W$ are the height and width of a frame, respectively. $C_{(i,j)}$ refers to the pixel values of the pixel in the frame at position $(i,j)$. $\bar{C}$ denotes the mean pixel value of the frame. We can adjust the position of the pixel value from $(i+1,j)$ to $(i,j+1)$ and $(i+1,j+1)$, to calculate the horizontal and diagonal correlation coefficients. The entropy can be calculated as follows:

$$H(m) = \sum_{i=0}^{2^N-1} P(m_i)\log_2 \frac{1}{P(m_i)} \text{ bits} \quad (7)$$

where $P(m_i)$ denotes the probability of pixel value $m_i$ and the entropy is presented in bits. The correlation coefficient and entropy of our method and other methods [58-62] are shown in Table 4. Although our video encryption method does not outperform the other image and video encryption methods, our method encodes the plain video into a neural network, with a relatively high degree of confusion and security in model paremeters, while adversaries cannot obtain the details of positional encoding method. From Table 5, to achieve a diverse array of encryption solutions, we have diligently crafted key-controllable modules with varying embedding lengths. These modules introduce enhanced flexibility and diversity into the encryption process, thereby enabling us to fulfill encryption requirements across a spectrum of security needs. By implementing such a design, we guarantee the robustness and security of our encryption algorithm, ultimately elevating the level of data protection achieved.

Table 4. Correlation coefficient of two adjacent pixels and entropies in plain and encrypted frames.

| Method | Pixels Correlation | | | Entropy |
|---|---|---|---|---|
| | Horizontal | Vertical | Diagonal | |
| Plain | 0.9897 | 0.9942 | 0.9819 | 7.625 |
| Ours | 0.8934 | 0.8512 | 0.7839 | 5.033 |
| SKK [58] | -0.0008 | 0.0003 | -0.0004 | 7.345 |
| Tanaka [59] | 0.2961 | 0.3616 | 0.2113 | 7.682 |
| Hybrid model [60] | -0.0006 | 0.0009 | -0.0006 | 7.999 |
| AES [61] | 0.0002 | -0.0002 | -0.0017 | 5.133 |
| Video encryption [62] | 0.0138 | -0.0321 | 0.0230 | 7.983 |

Table 5. PSNR and MS-SSIM results for different embedding lengths $2l$ of key-controllable module.

| Embedding length $2l$ | Key-controllable | PSNR | MS-SSIM |
|---|---|---|---|
| $2l=80$ | w/ key | 30.90 | 0.9440 |
| | w/o key | 5.34 | 0.1292 |
| $2l=120$ | w/ key | 30.91 | 0.9439 |
| | w/o key | 5.88 | 0.0918 |
| $2l=160$ | w/ key | 31.13 | 0.9462 |
| | w/o key | 5.85 | 0.0772 |

## C. Video Protection Efficiency

Video Protection efficiency measures the space and time efficiency of the decrypted video represented by the proposed model and tests the effect on space and time cost after the model is compressed and pruned. The enhanced quality of visual representation and model compression performance directly translates into reduced transmission and space costs. We evaluated the video content protection performance of our method under different model training epochs and parameters. Furthermore, we have quantified the cost of secure transmission by considering the size of model parameters, providing a comprehensive evaluation framework.

**Model compression techniques.** We used two model compression methods: model pruning and weight quantization, to reduce the number of parameters in NeR-VCP. Through model compression technology, we can deliver cipher videos at lower transmission costs and avoid suspicion from the adversary. For model pruning, we use global unstructured pruning in Torch to globally prune tensors corresponding to all parameters $\theta$:

$$\theta_i = \begin{cases} \theta_i, & \text{if } \theta_i \geq \theta_\partial \\ 0, & \text{otherwise} \end{cases} \tag{8}$$

Through Equation 8, set the parameter $\theta_\partial$ whose value is in the $\partial$ percentile bit to a threshold, which for $\theta_i$ is less than $\theta_\partial$ is set to zero. The pruned model was then fine-tuned and retrained for 100 epochs. For model quantization, we quantify the model at the end of the training process, which is different from some training methods based on semi-precision or mixed precision, which means that it is only quantized post-hoc:

$$\mu_i = \gamma\left(\frac{\mu_i - \mu_{min}}{2^{bit}}\right) * S + \mu_{min}, \quad S = \frac{\mu_{max} - \mu_{min}}{2^{bit}} \quad (9)$$

where $\gamma$ is the rounding value to the closest integer, $bit$ is the bit length for the quantized model, $\mu_{max}$ and $\mu_{min}$ are the max and min values for the parameter tensor $\mu$, respectively, and $S$ is the scaling factor. In Equation 9, each parameter in the model can be mapped to a $bit$ long value. Figure 10 (c & d) shows that the PSNR decreases to varying degrees after compression, but the basic decoding performance is maintained. In model pruning, when the degree of sparsity of the model reaches 0.2, PSNR decreases by 4, 6, and 8 dB. Moreover, the quantization bit is 4 bits, so the performance of the model will be greatly affected by weight quantization.

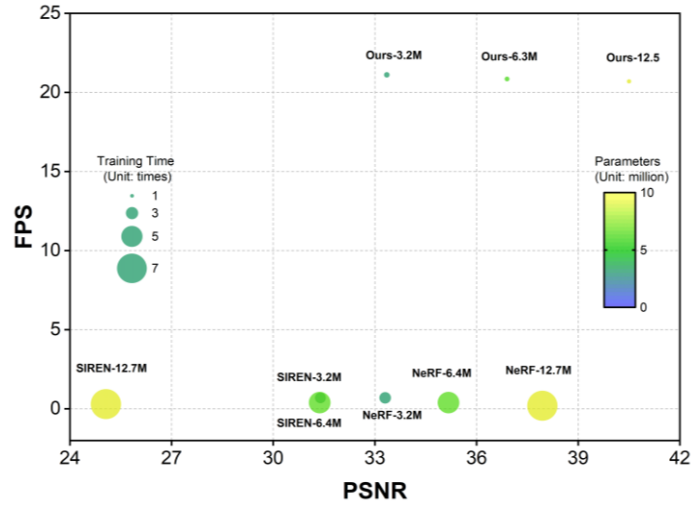

FIGURE 9. The model size, model performance, encoding FPS, and training time of different neural representation methods.

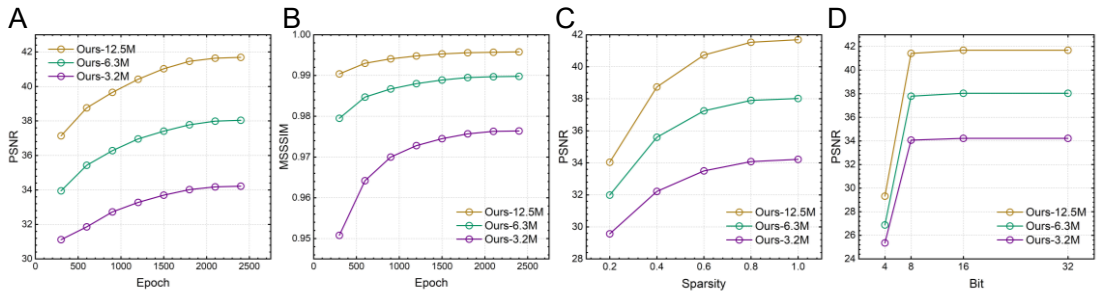

FIGURE 10. The visual performance with different epochs and model compression. A and B. The PSNR

and MS-SSIM results for different epochs. C and D. The experimental results with model quantization and model pruning.

**Performance on "Big Buck Bunny" video.** We compare our NeR-VCP with other implicit neural representation methods on "Big Buck Bunny" video. We chose SIREN [43] and NeRF [49] as baseline methods, and we changed the layer dimensions to introduce different sizes of models. For the proposed model, the NeR-VCP-3.2 M experiment on 'big buck bunny' can be reproduced with 26 for $C_1$, NeR-VCP-6.3 M and NeR-VCP-12.5 M with 58 and 112 for $C_1$, respectively, where $C_2/n$ is not less than 96. According to Figs. 9 and 10, for the same model size, we achieve better performances on both the PSNR and the decoded FPS than the other two methods. As the size of model increases, the PSNR of the decrypted video increases from 33.3 dB to 40.5 dB and the decoding FPS remains above 20. Second, given different timestamps, NeR-VCP can decode a series of frames in secret videos, so NeR-VCP takes less time than other methods to encode secret videos. In Fig. 10 (a & b), when we experimented with different training epochs, we found that longer training times can train better video overfitting results, and we note that as long as more epochs are trained, the final visual performance (PSNR and MS-SSIM) is better.

**Performance on the UVG dataset.** In addition, for the high efficiency of the proposed scheme, we also conducted a full test on the UVG dataset. In the experiment, we selected 7 videos from the UVG dataset, with a total of 3900 frames, as samples. The richer the video sequence in the time domain is, the better the video effect expressed by the model. We further tested the efficiency of NeR-VCP with video compression standards and other video representation methods: NeRV and E-NeRV. The framework of the model contains a large number of parameters such as weight and bias. Generally, the performance of a large model is better than that of other small models under the same structure. However, in the transmission process, we need to consider the transmission cost and other factors, which cannot increase the size of the model. Here, we use bits per pixel (BPP) as a calculation method to evaluate the efficiency of the representation. The expression of equation is as follows:

$$BPP = \frac{P_\theta \times Sparsity \times QB}{N_{pixel}} \qquad (10)$$

where $N_{pixel}$ is the number of pixels in the plain videos and $P_\theta$ is the number of parameters in the NeR-VCP network. $Sparsity$ is the sparsity of model, $QB$ indicates the bit length of the quantized parameter value.

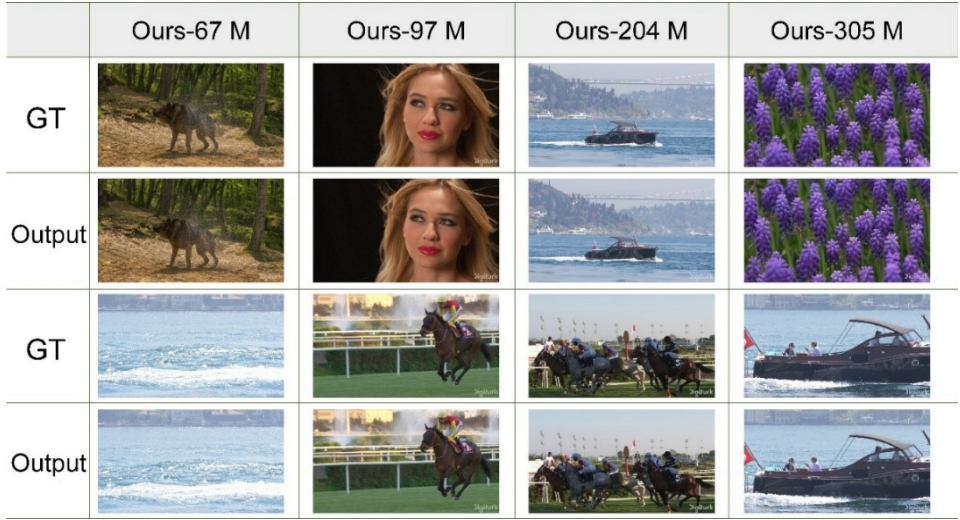

FIGURE 11. Visualization results for 1080p video dataset. NeR-VCP results are shown in experiment. Model sizes are indicated above the table.

Table 6. The parameters of different model sizes.

| | NeR-VCP-1080p | | | | | | |
|---|---|---|---|---|---|---|---|
| $C_1$ | 48 | 64 | 128 | 128 | 128 | 192 | 256 |
| $C_2$ | 384 | 512 | 512 | 768 | 1024 | 1536 | 2048 |
| Parameters | 18 M | 29 M | 46 M | 69 M | 97 M | 204 M | 350 M |

To improve the representation efficiency on the UVG dataset, we change the NeR-VCP size by adjusting the hyperparameters of $C_1$, $C_2$ to (48,384), (64,512), (128,512), (128,768), (128,1024), (192,1536), and (256,2048). As shown in Table 6 and Fig. 11, we provide the visual performance and size of different models with 300 epochs. To compare the performance of different sizes of model parameters, we selected two video compression algorithms, HEVC and H.264, and measured the impact of structure on the visual performance of decrypted videos from two metrics: PSNR and MS-SSIM. We use the default video encoding settings in FFmpeg and adjust the BPP with the constant rate factor value parameter. Figure 12 shows the experimental results of BPP *vs*. PSNR and MS-SSIM on UVG dataset with a sparsity of 1 and QB of 8. With MS-SSIM and PSNR as the metrics, although the neural representation methods for video and our method have not already achieved satisfactory performance compared to video compression standards, this is due to the lack of adequate training epochs. Neural representation methods have much better results for overfitting videos. For the proposed model, the final performances improve as the number of training epochs increases.

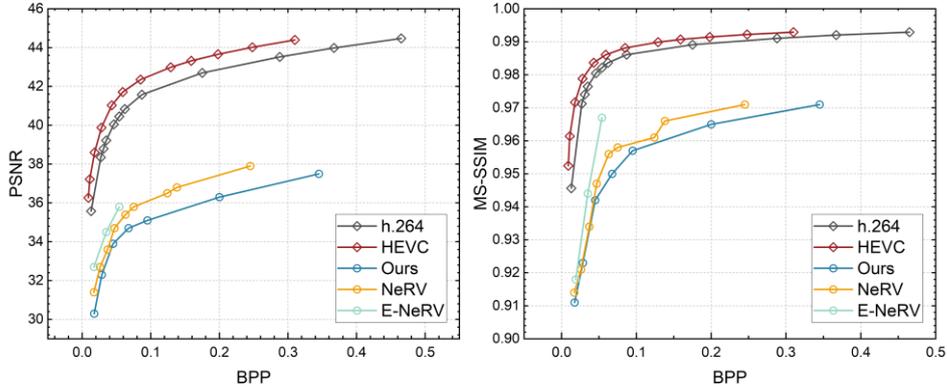

FIGURE 12. Comparison results of NeR-VCP and other state-of-the-art methods.

**Time Cost Analysis.** Time cost serves as an important metric for assessing the efficiency of video encryption method. This metric specifically pertains to the time required for encrypting and decrypting the video. We selected NeR-VCP 3.2 M with epoch of 300 to compare with other traditional and deep encryption methods [63-67]. As shown in Table 7, the encryption time of our proposed method exceeds that of other encryption methods. This is primarily attributed to the limited performance of data center GPU. It brings the prolonged time required for training model. However, the utilization of better deep learning GPU like A100 would significantly curtail the training time of our model by over 60%. Furthermore, our method boasts the capability to compress videos without the need for video compression standards, thereby mitigating the time of video coding.

Table 7. The time cost comparison between our method and other existing methods.

| Methods | Encrypted time/s | Decrypted time/s |
|---|---|---|
| Ours | 40.91 | 0.128 |
| Wu [63] | 31.45 | 35.66 |
| Ding [64] | 0.311 | 0.535 |
| Erkan [65] | 6.425 | 5.675 |
| Shariatzadeh [66] | 2.275 | 4.446 |
| Liao [67] | 3.241 | 3.448 |

## V  CONCLUSION

In this paper, we propose a video content protection method based on implicit neural representation. Following previous video encryption methods, our method combines video neural representation with video encryption to achieve content protection with video compression. We design a secure key-controllable scheme that boosts the performance of content protection with a good visual security. Using model compression techniques, our method with different parameters can perform well on video compression with state-of-the-art methods. The experiments show that our method has satisfactory effectiveness and imperceptibility from illegal access, and security in content protection.